\documentstyle[12pt,axodraw]{article}
\def\slashchar#1{\setbox0=\hbox{$#1$}           
   \dimen0=\wd0                                 
   \setbox1=\hbox{/} \dimen1=\wd1               
   \ifdim\dimen0>\dimen1                        
      \rlap{\hbox to \dimen0{\hfil/\hfil}}      
      #1                                        
   \else                                        
      \rlap{\hbox to \dimen1{\hfil$#1$\hfil}}   
      /                                         
   \fi}                                         %

\begin{document}
\begin{center}
{\Large\bf Two-particle decays of $B_c$ meson \\ into charmonium states}
\vspace{4mm}

{\sl  V.V.Kiselev}\\[1mm]
Russian State Research Center ``Institute for High Energy Physics'',\\
Protvino, Moscow Region, 142280 Russia\\[2mm]
and\\[2mm]
{\sl O.N.Pakhomova, V.A.Saleev}\\[2mm]
Samara State University, Samara, 443011 Russia
\end{center}

\begin{abstract}
The factorization of hard and soft contributions into the hadronic decays of
$B_c$ meson at large recoils is explored in order to evaluate the decay rates
into the  S, P and D-wave charmonia associated with $\rho$ and $\pi$. The
constraints of approach applicability and uncertainties of numerical estimates
are discussed. The mode with the $J/\psi$ in the final state is evaluated
taking into account the cascade radiative electromagnetic decays of excited
P-wave states, that enlarges the branching ratio by 20-25\%.
\end{abstract}

\section{Introduction}
After the first observation of $B_c$ meson by the CDF Collaboration at FNAL
\cite{cdf} in the semileptonic mode with the $J/\psi$ particle in the final
state,
\begin{equation}
B_c^+\to J/\psi  l^+\nu_l,
\label{1}
\end{equation}
one expects rather a significant, factor 20, increase of statistics with
$B_c$ in the same mode after RunII. However, the uncertainties in the mass
measurements are essential in the semileptonic channel, since the neutrino
momentum is not detected directly. So, the two-particle decay mode 
\begin{equation}
B_c^+\to J/\psi  \pi^+,
\label{2}
\end{equation}
is the most prospective channel for such the measurements. Therefore, we need a
qualitative theoretical modelling in order to predict the basic
characteristics of (\ref{2}). 

The dynamics of $B_c$ decays was studied in various theoretical approaches: the
QCD sum rules \cite{QCDSRBc} and potential models \cite{PMBc} operated with the
exclusive decays and gave the estimates for both the branching ratios and the
total lifetime summed over the exclusive modes, which is consistent with
estimates of inclusive decays and lifetime in the framework of Operator Product
Expansion (OPE) combined with the machinery of effective theory \cite{OPEBc} in
the form of nonrelativistic QCD (NRQCD) \cite{NRQCD}. However, the feature of
two-particle decays, which are studied in the present paper
\begin{equation}
B_c^+\to c\bar c [{\scriptstyle ^{2s+1}L_J}]  \pi^+(\rho^+),
\label{2a}
\end{equation}
is specified by rather a large recoil momentum of charmonium $c\bar c
[{\scriptstyle ^{2s+1}L_J}]$, where $s$ denotes the sum of quark and antiquark
spins, $L$ is the orbital quantum number running from 0 to 2, and $J$ is the
total spin of charmonium. Indeed, in the framework of potential models the
approximation of heavy-quarkonium wave-function overlapping for the calculation
of hadronic form factors can be explored in the region, where those
wave-functions are not exponentially small, i.e. if an amplitude under
consideration is soft enough, and the nonperturbative modelling in the form of
wave functions is reliable, while at large recoils the behaviour of vertices
for the quarks entering the bound states is significantly modified due to hard
gluon corrections. Then, the exponential decrease of quark-meson form factors
is replaced by the power-like one at large recoils. In that case, one could
factorize the hard and soft amplitudes \cite{Brodsky}, which was recently
explored in the description of two-particle decays of $B_c$ \cite{HSBc} as well
as heavy-light mesons \cite{Anisovich}.

While in \cite{HSBc} the decays into the S and P-wave charmonia were
considered, in the present paper we develop the same technique for a more
accurate analysis including the D-wave states of $c\bar c$, too. So, the
approach is based on the fact that in the heavy quarkonium we can neglect the
binding energy $\epsilon$ in comparison with the heavy quark mass, since by
the order of magnitude it is determined by the kinetic energy of heavy quark
and antiquark inside the meson $\epsilon \sim m_Q \cdot v^2$, where $v$ is the
relative velocity of quarks, $v\ll 1$, so that $\epsilon \ll m_Q$. Moreover,
the region of soft part in the heavy-quarkonium wave-function is determined by
the meson size, which is about $r\sim 1/p_Q$ with the quark momentum $p_Q\sim
m_Q\cdot v$. Then we can apply the nonrelativistic wave functions in the
amplitudes, where the virtualities $\mu^2$ are less than $(m_Q\cdot v)^2$,
while at large recoils in the decays the hard factors of amplitudes with
virtualities greater than $(m_Q\cdot v)^2$ should be described with account of
hard gluon exchange between the constituents of heavy quarkonia. The
perturbative QCD can be used for the hard amplitudes, if $\mu^2 \gg
\Lambda_{QCD}^2$. We check these conditions of hard-soft factorization and
estimate the uncertainties of numerical results by the variation of charmed
quark mass in the limits constrained by the excitation energy of P and D-levels
with respect to S-one.

In addition, we factorize the matrix element of light quark current by the
vacuum insertion. This approach and limits of its applicability are discussed
in \cite{fact}. Then, we deal with the hard approximation for the four
heavy-quark operator omitting possible renormalization effects. We take into
account the perturbative corrections to the effective nonleptonic weak
Lagrangian.

The paper is organized as follows: In Section 2 we present the basic model
assumptions and general formalism, while the analytic expressions for the
widths of $B_c$ decays into the charmonium states with the pion are given in
Section 3. We describe the input parameters and present the numerical results,
too. Then we take into account the radiative electromagnetic decays of P-levels
in order to estimate the summed $J/\psi \pi$ yield in the $B_c$ decays. In
Conclusion we discuss the obtained results and their uncertainties. Bulky
analytical expressions for the decays with $\rho$ are placed in the Appendix.

\section{Hard-soft factorization in $B_c$ decays}
Neglecting the binding energy, in the soft amplitude we put the mass of $B_c$
meson $m_1$ equal to the sum of b-quark and c-quark masses $m_b+m_c$, and the
mass of ${c\bar c}$ state $m_2$ equal to $2m_c$. Then the heavy quark and
antiquark inside the bound state move with the same 4-velocity, so that in the
accepted nonrelativistic approximation we can write down
\begin{eqnarray}
v_1 &=& \frac{p_1}{m_1}=\frac{p_{\bar b}}{m_b}=\frac{p_c}{m_c},
\label{s1}\\
v_2 &=& \frac{p_2}{m_2}=\frac{p_{\bar c}}{m_c}=\frac{p^{'}_c}{m_c},
\label{s2}
\end{eqnarray}
where $p_{1,2}$ are the momenta of decaying and recoil heavy quarkonia,
respectively, and $p_Q$ are the momenta of quarks composing the heavy
quarkonia.

However, at large recoils specific for the decays of $B_c^+\to c\bar c
[{\scriptstyle ^{2s+1}L_J}]  \pi^+(\rho^+)$, the conditions of (\ref{s1}) and
(\ref{s2}) could be valid only if we take into account the hard gluon
correction with a large momentum transfer
\begin{equation}
|k^2|=\frac{m_2}{4m_1}((m_1-m_2)^2-m_3^2)\gg\Lambda_{QCD}^2,
\end{equation}
where $m_3$ is the mass of $\pi$ or $\rho$. At the tree level as well as with
soft gluon corrections the prescription of (\ref{s1}) and (\ref{s2}) would give
zero matrix element, since a little smearing by soft gluons responsible for the
formation of wave-function results in the exponential suppression of
overlapping at large recoils. Numerically, the characteristic virtuality in the
hard amplitude is equal to $1.0-1.2$ GeV$^2$ for the charmonium in the final
state with $m_2=3.0-3.5$ GeV. We see that such virtualities are large enough
for quite a reliable use of perturbative QCD. Moreover, a characteristic
relative momentum of heavy quarks inside the bound states under consideration
is about $p\sim 0.6-0.7$ GeV, and the ratio $p^2/k^2 \sim 0.3-0.4$ is quite a
small parameter for the expansion. Thus, the kinematical conditions in the
decays of $B_c^+\to c\bar c [{\scriptstyle ^{2s+1}L_J}]  \pi^+(\rho^+)$ favor
the application of hard-soft approximation with the accuracy about 30\%.
Another source of uncertainty is connected with neglecting the binding energy,
and it is more essential. We will further test it numerically by the variation
of charmed quark mass from 1.5 to 1.7 GeV.

A general covariant formalism for calculating the production and decay rates of
S-wave and P-wave heavy quarkonium in the nonrelativistic expansion was
developed in \cite{n4}. In this approach, the amplitude for the decay
of bound state ($\bar b c$) possessing the momentum $p_1$, total spin $J_1$,
orbital momentum $L_1$ and summed spin $S_1$ into the bound-state ($\bar c c$)
possessing the momentum $p_2$, total spin $J_2$, orbital momentum $L_2$ and
summed spin $S_2$ is given by the following:
\begin{eqnarray}
A(p_1,p_2) &=&
\int \frac{d{\bf q}_1}{(2\pi)^3}\sum_{L_{1z}S_{1z}}
\Psi_{L_{1z}S_{1z}}({\bf q}_1)
\langle L_1L_{1z};S_1S_{1z}|J_1J_{1z}\rangle\times \\
&&\int \frac{d{\bf q}_2}{(2\pi)^3}\sum_{L_{2z}S_{2z}}
\Psi_{L_{2z}S_{2z}}({\bf q}_2) \langle L_2L_{2z};S_2S_{2z}|J_2J_{2z}\rangle 
M(p_1,p_2,q_1,q_2),\nonumber
\end{eqnarray}
where $M(p_1,p_2,q_1,q_2)$ is the hard amplitude of process with truncated
fermion legs entering the initial and final mesons as described by the diagrams
in Fig. \ref{fig1}. Here $\Psi_{L_zS_z}({\bf q})$ are the nonrelativistic wave
functions for the heavy quarkonia.

\begin{figure}
\begin{center}
\begin{picture}(250,100)(0,0)
\Text(10,40)[]{$p_1$}
\Line(20,43)(30,43)
\Line(20,37)(30,37)
\Line(33,40)(28,35)
\Line(33,40)(28,45)
\GOval(39,40)(35,6)(0){0.7}
\Oval(39,40)(35,6)(0)
\ArrowLine(168,75)(125,75)
\Text(105,67)[]{$u_1$}
\ArrowLine(125,75)(82,75)
\Line(82,75)(90,90)
\Line(82,75)(96,86)
\Line(96,91)(87,90)
\Line(96,91)(97,83)
\Text(98,97)[l]{$\pi^+(\rho^+)$}
\ArrowLine(82,75)(39,75)
\Gluon(125,75)(125,5){5}{5}\Text(118,40)[r]{$k$}
\ArrowLine(39,5)(82,5)
\ArrowLine(82,5)(125,5)
\ArrowLine(125,5)(168,5)
\GOval(168,40)(35,6)(0){0.7}
\Oval(168,40)(35,6)(0)
\Line(174,43)(184,43)
\Line(174,37)(184,37)
\Line(187,40)(182,35)
\Line(187,40)(182,45)
\Text(197,40)[]{$p_2$}
\end{picture}
\begin{picture}(200,100)(0,0)
\Text(10,40)[]{$p_1$}
\Line(20,43)(30,43)
\Line(20,37)(30,37)
\Line(33,40)(28,35)
\Line(33,40)(28,45)
\GOval(39,40)(35,6)(0){0.7}
\Oval(39,40)(35,6)(0)
\ArrowLine(168,75)(125,75)
\Text(105,67)[]{$u_2$}
\ArrowLine(125,75)(82,75)
\Line(125,75)(133,90)
\Line(125,75)(139,86)
\Line(139,91)(130,90)
\Line(139,91)(140,83)
\Text(141,97)[l]{$\pi^+(\rho^+)$}
\ArrowLine(82,75)(39,75)
\Gluon(82,75)(82,5){5}{5}\Text(75,40)[r]{$k$}
\ArrowLine(39,5)(82,5)
\ArrowLine(82,5)(125,5)
\ArrowLine(125,5)(168,5)
\GOval(168,40)(35,6)(0){0.7}
\Oval(168,40)(35,6)(0)
\Line(174,43)(184,43)
\Line(174,37)(184,37)
\Line(187,40)(182,35)
\Line(187,40)(182,45)
\Text(197,40)[]{$p_2$}
\end{picture}
\end{center}
\caption{The diagrams with the hard gluon exchange in decays of $B_c^+\to c\bar
c [{\scriptstyle ^{2s+1}L_J}]  \pi^+(\rho^+)$.}
\label{fig1}
\end{figure}
With the accuracy up to second order in relative momenta $q_1$ and $q_2$, the
operators $\Gamma_{SS_z}(p,q)$ projecting the quark-antiquark pairs onto the
bound states with fixed quantum numbers can be written in the form
\begin{equation}
\Gamma_{S_1S_{1z}}(p_1,q_1)=\frac{\sqrt{m_1}}{4m_cm_b}
(\frac{m_c}{m_1}\slashchar p_1-\slashchar q_1+m_c)\slashchar A_1
(\frac{m_b}{m_1}\slashchar p_1+\slashchar q_1-m_b),\label{8l}
\end{equation}
where $\slashchar A_1=\gamma_5$ for the pseudoscalar initial state $S_1=0$ and
$\slashchar A_1=\slashchar \varepsilon(S_{1z})$ for the vector one $S_1=1$, and
\begin{equation}
\Gamma^\dagger_{S_2S_{2z}}(p_2,q_2)=\frac{\sqrt{m_2}}{4m_c^2}
(\frac{m_c}{m_2}\slashchar p_2+\slashchar q_2-m_c)\slashchar A_2
(\frac{m_c}{m_2}\slashchar p_2-\slashchar q_2+m_c),\label{9l}
\end{equation}
where $\slashchar A_2=\gamma_5$ for the summed spin $S_2=0$ of recoil
quarkonium, and $\slashchar A_2=\slashchar \varepsilon(S_{2z})$ for $S_2=1$.
Here $\varepsilon(S_{1z,2z})$ denotes the polarization of vector-spin state.
The color factor $\delta^{ij}/\sqrt{3}$ for the singlet states should be also
introduced in the quark-meson vertices.

Making use of projection operators in (\ref{8l}) and (\ref{9l}) we write down
the hard amplitude $M(p_1,p_2,q_1,q_2)$ in the following way:
\begin{equation}
M(p_1,p_2,q_1,q_2)=\mbox{Tr}\left [
\Gamma^\dagger
(p_2,q_2)\gamma^{\mu}\Gamma(p_1,q_1)\cal{O}_{\mu}\right ],
\end{equation}
where for the decay with the $\pi$ meson in the final state we have
\begin{eqnarray}
{\cal O}_{\mu}&=&{\cal O}^{[1]}_{\mu}+{\cal O}^{[2]}_{\mu},\\
{\cal
O}^{[1]}_{\mu}&=&\frac{G_F}{\sqrt{2}}\frac{16\pi\alpha_s}{3}V_{bc}f_{\pi}a_1
\slashchar p_3 (1-\gamma_5)
\left (\frac{-\slashchar u_1+m_c}{x_1^2-m_c^2}
\right )\frac{\gamma_{\mu}}{k^2},\\
{\cal
O}^{[2]}_{\mu}&=&\frac{G_F}{\sqrt{2}}\frac{16\pi\alpha_s}{3}V_{bc}f_{\pi}a_1
\frac{\gamma_{\mu}}{k^2}\left (\frac{-\slashchar u_2+m_b}{x_2^2-m_b^2}
\right )\slashchar p_3(1-\gamma_5),
\end{eqnarray}
with the notations
$$ 
\slashchar u_1=\frac{m_b}{m_1}\slashchar p_1+\slashchar q_1-\slashchar p_3,
\quad \slashchar u_2=\frac{m_c}{m_2}\slashchar p_2+\slashchar q_2+\slashchar
p_3, \quad
 \slashchar k=\frac{m_c}{m_2}\slashchar p_2-\frac{m_c}{m_1}\slashchar
p_1+\slashchar q_1-
\slashchar q_2.$$

The factor $a_1$ comes from the hard gluon corrections to the four-fermion
effective weak Lagrangian. Expanding $M(p_1,p_2,q_1,q_2)$ over small parameters
$q_1/m_1$ and $q_2/m_2$ at $q_1=q_2=0$, we get
\begin{eqnarray}
M(p_1,p_2,q_1,q_2) &=& M(p_1,p_2,0,0)+
q_{1\alpha}\frac{\partial M}{\partial q_{1\alpha}}|_{q_{1,2}=0}+
q_{2\alpha}\frac{\partial M}{\partial
q_{2\alpha}}|_{q_{1,2}=0}+\nonumber\\
&+&\frac{1}{2}q_{2\alpha}q_{2\beta}
\frac{\partial^2 M}{\partial q_{2\alpha}
\partial q_{2\beta}}|_{q_{1,2}=0}+\ldots
\end{eqnarray}
where the above terms correspond to various quantum numbers, first, with 
$L_1=L_2=0$ for the transition between the S-wave levels, second and third,
with $L_1=1$ and $L_2=0$, $L_1=0$ and $L_2=1$ for the transitions between S and
P-wave states, fourth, with $L_1=0$ and $L_2=2$ for the transition between the
S-wave initial quarkonium and the D-wave recoil meson. Thus, we can easily find
that for the various orbital states, the soft factors in the amplitudes
$A(p_1,p_2)$ are expressed in terms of  quarkonium radial wave-functions in the
following way:
\begin{eqnarray}
\int \frac{ d^3 {\bf q}}{(2\pi)^3} \; \Psi_{00} ({\bf q} )
&=& \frac{R(0)}{\sqrt{4\pi}} \; , \nonumber \\
\int \frac{ d^3 {\bf q}}{(2\pi)^3} \; \Psi_{1L_Z} ({\bf q} ) q_\alpha
&=& - i \sqrt{\frac{3}{4\pi}} R'(0) \varepsilon_\alpha(p,L_Z) \; ,
\nonumber \\
\int \frac{d^3{\bf q}}{(2\pi)^3} \; \Psi_{2L_Z} ({\bf q} ) q_\alpha q_\beta
&=& \sqrt{\frac{15}{8\pi}} R''(0) \varepsilon_{\alpha\beta}(p,L_Z)
\; ,
\end{eqnarray}
where $\varepsilon_{\alpha}(p,L_Z)$ is the polarization of vector particle, and
$\varepsilon_{\alpha\beta}$ is the symmetric, traceless, and transverse rank-2
polarization for the spin-2 particle. The above wave-functions are represented
as the products of radial and angular functions: $\Psi_{L L_Z}({\bf r}) = Y_{L
L_Z}(\theta,\phi)\, R_{L}(r)$.

For the $^1P_1$ charmonium state we get
\begin{equation}
\sum_{L_{2Z}}\varepsilon^{\alpha}(p_2,L_{2Z})\langle 1L_{2Z},00\vert
1,J_{2Z}\rangle =\varepsilon^{\alpha}(p_2,J_{2Z}).
\end{equation}
For $^3P_J(J=0,1,2)$ states the summation over the quark spins and orbital
momentum projections results in 
\begin{eqnarray}
\sum_{S_{2Z},L_{2Z}}\varepsilon^{\alpha}(p_2,L_{2Z})\langle 1L_{2Z},1S_{2Z}
\vert J_2,J_{2Z}\rangle \varepsilon^{\rho}(S_{2Z})
=\left \{ \begin{array}{lr}
\frac{1}{\sqrt{3}}(g^{\alpha\rho}-\frac{p_2^{\alpha}p_2^{\rho}}
{m_2^2}),& \mbox{ }J_2=0,\\[2mm]
\frac{i}{\sqrt{2}m_2}\varepsilon^{\alpha\rho\mu\nu}p_{2\mu}
\varepsilon_{\nu}(p_2,J_{2Z}),&\mbox{ }J_{2}=1,\\[2mm]
\varepsilon^{\rho\alpha}(p_2,J_{2Z}),&\mbox{ } J_2=2.
\end{array}
\right.
\end{eqnarray}
For $c\bar c[{\scriptstyle ^1D_2}]$ one has
\begin{equation}
\sum_{L_{2Z}}\varepsilon^{\alpha\beta}(p_2,L_{2Z})\langle 2L_{2Z},00\vert
2,J_{2Z}\rangle =\varepsilon^{\alpha\beta}(p_2,J_{2Z}).
\end{equation}
For $c\bar c[{\scriptstyle ^3D_J(J=1,2,3)}]$ states we get
\begin{eqnarray}
&&\sum_{S_{2Z},L_{2Z}}\varepsilon^{\alpha\beta}(p_2,L_{2Z})\langle
2L_{2Z},1S_{2Z} \vert J_2,J_{2Z}\rangle \varepsilon^{\rho}(S_{2Z})=\nonumber\\
&\quad &=\left \{ \begin{array}{lr}
  - \sqrt{\frac{3}{20}} \; \left(
\frac{2}{3} {\cal P}_{\alpha\beta} \; \varepsilon_\rho (p_2,J_{2Z})
-{\cal P}_{\alpha\rho}\;\varepsilon_\beta(p_2,J_{2Z}) - {\cal P}_{\beta\rho}
\;\varepsilon_\alpha(p_2,J_{2Z}) \right), &
\mbox{ }J_2=1,\\[2mm]
\frac{i}{M\sqrt{6}}\left( \varepsilon_{\alpha\sigma}(p_2,J_{2Z})
\varepsilon_{\tau\beta\rho  \sigma'}\; p_2^\tau g^{\sigma\sigma'} +
\varepsilon_{\beta\sigma}(p_2,J_{2Z})
\varepsilon_{\tau\alpha\rho\sigma'}\; p_2^\tau g^{\sigma\sigma'} \right),&
\mbox{ } J_2=2,\\[2mm]
\varepsilon_{\alpha\beta\rho} (p_2,J_{2Z}), &
\mbox{ } J_2=3,
\end{array}
\right.
\end{eqnarray}
where
\begin{equation}
{\cal P}^{\alpha\beta} = -g^{\alpha\beta} + \frac{p_2^\alpha
p_2^\beta}{m_2^2}\;,
\end{equation}
and $\epsilon_{\alpha\beta\rho}(p_2,J_{2Z})$ is the symmetric, traceless, and
transverse spin-3 polarization tensor. The summation over the polarizations
gives the following ordinary expressions \cite{n5}:
\begin{eqnarray}
\label{j=1}
\sum_{J_{2Z}=-1}^1  \varepsilon_\alpha(p_2,J_{2Z})
\varepsilon_\beta^*(p_2,J_{2Z}) &=&
       {\cal P}_{\alpha\beta} , \\
\label{j=2}
\sum_{J_{2Z}=-2}^2  \varepsilon_{\alpha\beta}(p_2,J_{2Z})
\varepsilon_{\rho\sigma}^*(p_2,J_{2Z}) &=&
\frac{1}{2}\left( {\cal P}_{\alpha\rho}
{\cal P}_{\beta\sigma}
+ {\cal P}_{\alpha\sigma} {\cal P}_{\beta\rho}
\right ) -\frac{1}{3} {\cal P}_{\alpha\beta}
{\cal P}_{\rho\sigma} \; , \\
\label{j=3}
\sum_{J_{2Z}=-3}^3  \varepsilon_{\alpha\beta\gamma}(p_2,J_{2Z})
                 \varepsilon_{\rho\sigma\eta}^*(p_2,J_{2Z}) &=&
\frac{1}{6} \biggr({\cal P}_{\alpha\rho} {\cal P}_{\beta\sigma}
{\cal P}_{\gamma\eta} + {\cal P}_{\alpha\rho} {\cal P}_{\beta\eta}
{\cal P}_{\gamma\sigma} + {\cal P}_{\alpha\sigma} {\cal P}_{\beta\rho}
{\cal P}_{\gamma\eta}
\nonumber \\
&& + {\cal P}_{\alpha\sigma} {\cal P}_{\beta\eta}
{\cal P}_{\gamma\rho} + {\cal P}_{\alpha\eta} {\cal P}_{\beta\sigma}
{\cal P}_{\gamma\rho} + {\cal P}_{\alpha\eta} {\cal P}_{\beta\rho}
{\cal P}_{\gamma\sigma} \biggr )
\\
&& -\frac{1}{15} \biggr({\cal P}_{\alpha\beta} {\cal P}_{\gamma\eta}
{\cal P}_{\rho\sigma} + {\cal P}_{\alpha\beta} {\cal P}_{\gamma\sigma}
{\cal P}_{\rho\eta} + {\cal P}_{\alpha\beta} {\cal P}_{\gamma\rho}
{\cal P}_{\sigma\eta}
\nonumber \\
&& + {\cal P}_{\alpha\gamma} {\cal P}_{\beta\eta}
{\cal P}_{\rho\sigma} + {\cal P}_{\alpha\gamma} {\cal P}_{\beta\sigma}
{\cal P}_{\rho\eta} + {\cal P}_{\alpha\gamma} {\cal P}_{\beta\rho}
{\cal P}_{\sigma\eta}
\nonumber \\
&& + {\cal P}_{\beta\gamma} {\cal P}_{\alpha\eta}
{\cal P}_{\rho\sigma} + {\cal P}_{\beta\gamma} {\cal P}_{\alpha\sigma}
{\cal P}_{\rho\eta} + {\cal P}_{\beta\gamma} {\cal P}_{\alpha\rho}
{\cal P}_{\sigma\eta} \biggr). \nonumber
\end{eqnarray}
To the moment, we completely define the procedure for the calculation of matrix
elements squared, that we perform by the use of analytic calculation system
MATHEMATICA.

\section{Results}
Neglecting the $\pi$ meson mass, we obtain the following analytical formulae
for the widths of $B_c^+\to c\bar c [{\scriptstyle ^{2s+1}L_J}] \pi^+(\rho^+)$:
\begin{eqnarray}
&&\Gamma (B_c\to \psi\pi)=\frac{128}{9\pi}F\frac{|R_2(0)|^2}{m_2^3}
\frac{(1+x)^3}{(1-x)^5},\\
&&\Gamma (B_c\to \eta_c\pi)=\frac{32}{9\pi}F\frac{|R_2(0)|^2}{m_2^3}
\frac{(1+x)^3}{(1-x)^5}(x^2-2x+3)^2,\\
&&\Gamma (B_c\to h_c\pi)=\frac{128}{3\pi}F\frac{|R'_2(0)|^2}{m_2^5}
\frac{(1+x)^3}{(1-x)^7}(x^3-2x^2+3x+4)^2,\\
&&\Gamma (B_c\to
\chi_{c0}\pi)=\frac{128}{\pi}F\frac{|R'_2(0)|^2}{m_2^5}
\frac{(1+x)^3}{(1-x)^7}(x^2-2x+3)^2,\\
&&\Gamma (B_c\to
\chi_{c1}\pi)=\frac{256}{3\pi}F\frac{|R'_2(0)|^2}{m_2^5}
\frac{(1+x)^3}{(1-x)^3},\\
&&\Gamma (B_c\to
\chi_{c2}\pi)=\frac{256}{\pi}F\frac{|R'_2(0)|^2}{m_2^5}
\frac{(1+x)^5}{(1-x)^7},\\
&&\Gamma (B_c\to
{^1D_2}\pi)=\frac{2560}{9\pi}F\frac{|R''_2(0)|^2}{m_2^7}
\frac{(1+x)^5}{(1-x)^9}(x^3-3x+5x+5)^2,\\
&&\Gamma (B_c\to
{^3D_1}\pi)=\frac{256}{9\pi}F\frac{|R''_2(0)|^2}{m_2^7}
\frac{(1+x)^3}{(1-x)^9}(5x^3-22x^2+41x+8)^2,\\
&&\Gamma (B_c\to
{^3D_2}\pi)=\frac{5120}{3\pi}F\frac{|R''_2(0)|^2}{m_2^7}
\frac{(1+x)^5}{(1-x)^5},\\
&&\Gamma (B_c\to
{^3D_3}\pi)=\frac{8192}{3\pi}F\frac{|R''_2(0)|^2}{m_2^7}
\frac{(1+x)^7}{(1-x)^9},
\end{eqnarray}
where
$$
x=\frac{m_2}{m_1}, \mbox{ and }
F=\alpha_s^2G_F^2V_{bc}^2f_{\pi}^2|R_1(0)|^2a_1^2.
$$
In numerical estimations we use the following set
of parameters: 
\begin{center}
\begin{tabular}{rclrclrcl}
$|R_{1}(0)|^2$ &=& $1.27$  GeV$^3$, \hspace*{3mm}&
$m_c$ &=& $1.5 $  GeV,\hspace*{3mm}& 
$m_{\pi}$ &=& $0.14$  GeV,\\
$|R_{2}(0)|^2$ &=& $0.94$ GeV$^3$,&
$m_b$ &=& $4.8 $  GeV,&
$f_{\pi}$ &=&$ 0.13$  GeV, \\
$|R'_{2}(0)|^2$ &=& $0.08$ GeV$^5$,&
$m_1$ &=& $6.3 $  GeV,&
$V_{bc}$ &=& $0.04$,\\
$|R''_{2}(0)|^2$ &=& $0.015$ GeV$^7.$&
$m_2$ &=& $3.0$ GeV,&
$\alpha_s$ &=& $0.33$.\\
\end{tabular}
\end{center}
The values of wave-function parameters are taken from \cite{EichQuigg}.
Then we get the estimate of direct $J/\psi$ yield associated with the pion 
\begin{equation}
    \Gamma (B_c^+\to J/\psi  \pi^+)=6.455 \times 10^{-15}a_1^2\mbox{ GeV}.
\end{equation}
The decay widths into different charmonium states and the $\pi$ meson
are presented in Table \ref{tab1} as the fractions of decay width for $B_c^+\to
J/\psi \pi^+$, while the absolute values depending on the choice of $a_1$ are
given in Table \ref{tab2}.

\begin{table}[th]
\caption{The yields of charmonium states in hadronic two-particle decays of
$B_c$ meson represented as the ratios.}
\label{tab1}
\begin{center}
\begin{tabular}{|c|c|c|c|}\hline &&&\\[-3mm]
${c\bar c}$ &
$^{2S+1}L_J$ &
$\displaystyle{\Gamma (B_c\to {c\bar c}[{\scriptstyle
^{2S+1}L_J}]\pi)\over\Gamma (B_c\to J/\psi\pi)} $&
$\displaystyle{\Gamma (B_c\to {c\bar c}[{\scriptstyle
^{2S+1}L_J}]\rho)\over\Gamma (B_c\to {c\bar
c}[{\scriptstyle ^{2S+1}L_J}]\pi)}$\\[4mm]
\hline
$J/\psi$ & $^3S_1$ & 1.0 & 3.9 \\[0.75mm] 
$\eta_c$ & $^1S_0$ & 1.3 & 3.2 \\[0.75mm] 
$h_c$    & $^1P_1$ & 2.7 & 3.4 \\[0.75mm] 
$\chi_{c0}$ & $^3P_0$& 1.6 & 3.5 \\[0.75mm] 
$\chi_{c1}$ & $^3P_1$& 0.016 & 51 \\[0.75mm] 
$\chi_{c2}$ & $^3P_2$& 1.4 & 4.0 \\[0.75mm] 
            & $^1D_2$& 5.4 & 3.6 \\[0.75mm] 
            & $^3D_1$& 2.8 & 3.8 \\[0.75mm] 
            & $^3D_2$& 0.053 & 31 \\[0.75mm] 
            & $^3D_3$& 2.4 & 4.2 \\[0.75mm]  \hline
\end{tabular}
\end{center}
\end{table}

\begin{table}[th]
\caption{The widths of $B_c$ meson with the charmonium states in hadronic
two-particle decays calculated in the hard-soft factorization in comparison
with the results of wave-function overlapping technique \cite{chache,10}.}
\label{tab2}
\begin{center}
\begin{tabular}{|c|c|p{2.5cm}|p{2.5cm}|p{2.5cm}|p{2.5cm}|}\hline
&&\multicolumn{2}{|c|}{}&
\multicolumn{2}{|c|}{}\\[-3mm]
${c\bar c}$ &
$^{2S+1}L_J$ & \multicolumn{2}{c}{
$\Gamma (B_c\to {c\bar c}[{\scriptstyle ^{2S+1}L_J}]\pi)$, $10^{-15}$ GeV} & 
\multicolumn{2}{|c|}{
$\Gamma (B_c\to {c\bar c}[{\scriptstyle ^{2S+1}L_J}]\rho)$, $10^{-15}$
GeV}\\[2mm]
\hline
$\eta_c$&${^1S_0}$&8.4 $a_1^2$ & 2.1 $a_1^2$ \cite{10} & 27 $a_1^2$ &  5.5
$a_1^2$ \cite{10}\\
$J/\psi$&${^3S_1}$&6.5 $a_1^2$ &  2.0 $a_1^2$ \cite{10} & 26 $a_1^2$ &  6.0
$a_1^2$ \cite{10}\\[0.75mm] 
$h_c$ &${^1P_1}$&18 $a_1^2$  & 0.57  $a_1^2$ \cite{chache}& 60 $a_1^2$ & 1.4
$a_1^2$ \cite{chache}\\[0.75mm] 
$\chi_{c0}$&${^3P_0}$ &11 $a_1^2$& 0.32 $a_1^2$ \cite{chache} & 37 $a_1^2$ &
0.81 $a_1^2$ \cite{chache}\\[0.75mm] 
$\chi_{c1}$ &${^3P_1}$&0.10 $a_1^2$ & 0.082 $a_1^2$ \cite{chache} & 5.2 $a_1^2$
& 0.33 $a_1^2$ \cite{chache}\\[0.75mm] 
$\chi_{c2}$ &${^3P_2}$&8.9 $a_1^2$ & 0.28 $a_1^2$ \cite{chache} & 36 $a_1^2$ &
0.58 $a_1^2$ \cite{chache}\\[0.75mm] 
&${^1D_2}$ &35 $a_1^2$ & & 124 $a_1^2$ & ~\\[0.75mm] 
&${^3D_1}$ &19 $a_1^2$ & & 70 $a_1^2$ & ~\\[0.75mm] 
&${^3D_2}$ &0.34 $a_1^2$ & & 11 $a_1^2$ & ~\\[0.75mm] 
&${^3D_3}$ &16 $a_1^2$ & & 65 $a_1^2$ & ~\\[0.75mm] 
\hline
\end{tabular}
\end{center}
\end{table}

An additional source of $J/\psi$ mesons is the two-particle decay of $B_c$
meson with $\rho$ in the final state: $B_c^+\to {c\bar c}[{\scriptstyle
^{2S+1}L_J}]\rho^+$. Calculating the decay widths $\Gamma (B_c^+\to {c\bar
c}[{\scriptstyle ^{2S+1}L_J}]\rho^+)$ can be done in the same way as for the
widths $\Gamma (B_c^+\to {c\bar c}[{\scriptstyle ^{2S+1}L_J}]\pi^+)$. Indeed,
we use the factorization of light meson current, so that in decays with $\rho$
we incorporate the substitution $f_{\pi} p_3^\mu\to m_{\rho} f_{\rho}
\varepsilon_3^\mu$, where $\varepsilon_3^{\mu}$ is the $\rho$ meson
polarization. Taking into account the numerical values of $f_{\rho}=0.22$ GeV
and $m_{\rho}=0.77$ GeV, we get the decay widths $\Gamma (B_c^+\to {c\bar
c}[{\scriptstyle ^{2S+1}L_J}]\rho^+)$, which are presented in Tables
\ref{tab1}, \ref{tab2}. Cumbersome analytical expressions for the decay widths
of $B_c^+\to {c\bar c}[{\scriptstyle ^{2S+1}L_J}]\rho^+$ are given in Appendix.

For the sake of comparison with the results obtained in the ordinary technique
with the wave-function overlapping, in Table \ref{tab2} we show also the
estimates, which were recently evaluated in \cite{chache,10}. The analysis of
two-particle hadronic $B_c$ decays with the charmed S-wave recoil-mesons in the
final state was also done in \cite{Vary}, where the estimates are similar with
those of \cite{10}, but slightly less numbers. Then, we see that our estimates
with the charmed quark mass fixed at $m_c=1.5$ GeV are significantly greater
than the values calculated in \cite{chache,10}. Indeed, for the S-wave
charmonium the increase due to the nonexponential fall off the wave-functions
is about a factor 4 in the matrix element squared, while for the P-waves this
factor reaches one order of magnitude. The reason for such the increase is
quite transparent. So, following the coulomb analogy, we can expect that the
velocity of heavy quark motion in the P-wave quarkonium is less than the
velocity in the S-wave state (remember, $v_n \sim \alpha/n$, where $n$ is the
principal quantum number). Then, the wave functions of P-wave levels have more
soft behaviour than in the S-wave states, i.e. the relative momentum of heavy
quarks is softer in the P-wave states, while the overlapping of quarkonia
wave-functions at large recoils is displaced into the region of high
virtualities, and the suppression is stronger for the P-wave levels. Thus, we
would expect the above result on the significant enhancement of P-wave level
yield in the hard-soft factorization approach. 

Another problem concerning for the factorization applicability is related to
inherent uncertainties due to neglecting the binding energy in the charmonium
states. Indeed, putting $m_c =m_{J/\psi}/2$ leads to zero binding energy for
the S-wave states, while for the excitations under study it is about 500 MeV,
i.e. it could be quite essential in the numerical estimates. We test this
dependence on the binding energy by the variation of charmed quark mass in the
range of $1.5-1.7$ GeV. We find that this variation brings the significant
uncertainties into the absolute values of widths under consideration, so that
the variations are about 30-50\% for the P-wave charmonia and greater than
100\% for the D-waves. Nevertheless, we observe that the ratios of widths
presented in Table \ref{tab1} are quite stable under such the variation of
charmed quark mass. The corresponding uncertainties are limited by $5-10$\%.
This fact implies that the theoretical predictions for the ratios of
two-particle widths in the $B_c$ decays are quite reliable. Moreover, these
ratios are close to the values obtained in \cite{chache,10}.

Since the $J/\psi$ meson is experimentally detected with a good efficiency in
the decays of $B_c$, we compare the direct $J/\psi$ yield ($B_c^+\to J/\psi
\pi^+(\rho^+)$) with the cascade one in decays with the radiative
electromagnetic transitions of excited P-wave states into $J/\psi$. The
corresponding branching ratios of radiative decays are known experimentally
\cite{n6}. So,
$$
\mbox{Br}(\chi_{c0}\to J/\psi\gamma)=0.007,\;
\mbox{Br}(\chi_{c1}\to J/\psi\gamma)=0.27, \;
\mbox{Br}(\chi_{c2}\to J/\psi\gamma)=0.14.
$$
Then we obtain
\begin{equation}
{\Gamma (B_c^+\to\chi_{c0,c1,c2}\pi^+\to J/\psi\pi^+\gamma)\over
{\Gamma (B_c^+\to J/\psi\pi^+)}}=0.21,
\end{equation}
and
\begin{equation}
{\Gamma (B_c^+\to\chi_{c0,c1,c2}\rho^+\to J/\psi\rho^+\gamma)\over
{\Gamma (B_c^+\to J/\psi\rho^+)}}=0.26.
\end{equation}
Thus, we see that the correction to the $J/\psi$ yield in the hadronic
two-particle decays of $B_c$ due to the indirect mechanism with the P-wave
charmonium is about 20-25\%. In contrast, analogous contribution in the
semileptonic decays of $B_c$ is significantly less, and the corresponding
fraction due to the P-wave charmonium is about 5\% \cite{PMBc}.

Supposing $a_1=1.1$, we get 
\begin{equation}
{\rm Br}(B_c^+\to J/\psi\pi^+)+{\rm Br}(B_c^+\to J/\psi\rho^+)\approx 2.8\%,
\end{equation}
and in the $B_c$ decays to $J/\psi$ with radiative transitions of
$\chi_{c0,c1,c2}$ the correction is equal to
\begin{equation}
{\rm Br}(B_c^+\to J/\psi\pi^+\gamma)+{\rm Br}(B_c^+\to
J/\psi\rho^+\gamma)=0.64\%.
\end{equation}

\section{Conclusion}

In this paper we have considered the hadronic two-particle decays of $B_c$
meson with large recoils in the technique of hard-soft factorization for the
matrix elements. This factorization is based on the physical separation of hard
rescattering the constituents composing the heavy quarkonia (the scale of
virtualities $\sim 1-1.5$ GeV$^2$) from the soft binding of heavy quarks (the
scale of virtualities $\sim 0.3-0.45$ GeV$^2$). Hard factors can be calculated
in the perturbative QCD, while the soft ones are expressed in terms of
wave-functions and their derivatives at the origin for S, P and D-wave levels
of heavy quarkonia. We have calculated the widths of $B_c^+\to {c\bar
c}[{\scriptstyle ^{2S+1}L_J}]\pi^+(\rho^+)$ decays for the charmonium in the
final states (see Tables \ref{tab1}, \ref{tab2}). We have found that the
results for the ratios of widths are quite stable under the variation of model
parameters, while the absolute values have rather large uncertainties
especially because of variation of charmed quark mass, that reflects the main
systematic uncertainty due to the approximation of zero binding energy in the
charmonium. We have compared our results with the potential model
\cite{chache,10} operating with the wave-function overlapping. The relative
momentum of charmed quarks inside the charmonium states (especially inside the
excited P and D-waves, where the relative velocity of heavy quarks becomes less
than for the S-levels) is rather small in comparison with the recoil momentum,
so that the overlapping is displaced into the exponentially suppressed region,
where the wave-function formalism is not reliable. In this region the hard
gluon corrections replacing the exponential behaviour of quark-meson form
factors by the power one, are significant. Thus, we expect valuable yields
of excited charmonium states in the two-particle decays of $B_c$. This increase
results in the additional 20-25\% fraction of $J/\psi\pi$ inclusive branching
ratio due to the contribution caused by the radiative electromagnetic
transitions of P-wave charmonium states.

The authors thank prof. A.K.Likhoded for fruitful discussions and valuable
remarks. This work is in part supported by the Russian Foundation for Basic
Research, grants 01-02-99315, 01-02-16585 and 00-15-96645, the Federal
program ``State support for the integration of high education and fundamental
science'', grant 247 (V.V.K. and O.N.P.), and the Federal program ``University
of Russia --- Basic Researches", grant 02.01.03.

\section*{Appendix}
In this appendix we write down cumbersome formulae for the ratios of decay
widths with the $\rho$ meson with respect to the widths with the pion for
various recoil charmonium sates. We introduce $x=m_2/m_1$, where $m_{1,2,3}$
are the masses of $B_c$, $c\bar c[{\scriptstyle ^{2s+1}L_J}]$ and $\rho$,
respectively. Then we get \small
\begin{eqnarray*}
&&\frac{\Gamma(B_c\to\eta_c\rho)}{\Gamma(B_c\to\eta_c\pi)}=
B_{\eta_c}\,\left(1-\frac{m_3^2}{m_1^2}\,
\frac{5x^2+2x+3}{x^2\,(1+x)^2\,(3x^2-2x+1)}\right)+{\cal
O}\left(\frac{m_3^4}{m_1^4}\right),\\[3mm]
&&\frac{\Gamma(B_c\to\ J/\psi\rho)}{\Gamma(B_c\to\ J/\psi\pi)}=
B_{J/\psi}\,\left(1+\frac{m_3^2}{2
m_1^2}\frac{2x^6+20x^5-3x^4-16x^3+16x^2-4x+1}
{x^6\,(1-x)^2\,(2+x)^2}\right)+{\cal O}\left(\frac{m_3^4}{m_1^4}\right),\\[3mm]
&&\frac{\Gamma(B_c\to\ h_c\rho)}{\Gamma(B_c\to\ h_c\pi)}=
B_{h_c}\, \left(1-\frac{2
m_3^2}{m_1^2}\,\frac{2x^6-3x^5+6x^4+8x^3+6x^2+31x+18}
{(1+x)^2\,(x^3-2x^2+3x+4)^2}\right)+
{\cal O}\left(\frac{m_3^4}{m_1^4}\right),\\[3mm]
&&\frac{\Gamma(B_c\to \chi_{c0}\rho)}{\Gamma(B_c\to \chi_{c0}\pi)}=
B_{\chi_{c0}}\,\left(1-\frac{m_3^2}{3 m_1^2}\,\frac{(2x^4-17x^3-5x^2+7x+29)}
{(1-x)\,(1+x)^2\,(x^2-2x+3)}\right)+{\cal
O}\left(\frac{m_3^4}{m_1^4}\right),\\[3mm]
&&\frac{\Gamma(B_c\to \chi_{c1}\rho)}{\Gamma(B_c\to \chi_{c1}\pi)}=
B_{\chi_{c1}}\, \Biggl(1+\frac{m_3^2}{2 m_1^2}\,
\frac{2}{(1-x)^6\,(1+x)^2}
(5x^8-24x^7+91x^6\\&&\hspace*{3cm}-158x^5+183x^4-4x^3-119x^2+130x+40)\Biggr)+
{\cal O}\left(\frac{m_3^4}{m_1^4}\right),\\[3mm]
&&\frac{\Gamma(B_c\to \chi_{c2}\rho)}{\Gamma(B_c\to \chi_{c2}\pi)}=
B_{\chi_{c2}}\,
\left(1+\frac{m_3^2}{6 m_1^2}\,\frac{x^6-6x^5+32x^4-48x^3-11x^2+78x-10}
{(1-x)^2\,(1+x)^2}\right)+{\cal O}\left(\frac{m_3^4}{m_1^4}\right),\\[3mm]
&&\frac{\Gamma(B_c\to c\bar c[{\scriptstyle ^1D_2}]\rho)}
{\Gamma(B_c\to c\bar c[{\scriptstyle ^1D_2}]\pi)}=
B_{c\bar c[{\scriptstyle ^1D_2}]}\times\\
&&\hspace*{3.7cm}\left(1-\frac{2
m_3^2}{m_1^2}\,\frac{3x^6-12x^5+27x^4-16x^3-3x^2+100x+49}
{(1+x)^2\,(x^3-3x^2+5x+5)^2}\right)+{\cal
O}\left(\frac{m_3^4}{m_1^4}\right),\\[3mm]
&&\frac{\Gamma(B_c\to c\bar c[{\scriptstyle ^3D_1}]\rho)}
{\Gamma(B_c\to c\bar c[{\scriptstyle ^3D_1}]\pi)}=B_{c\bar c[{\scriptstyle
^3D_1}]}\times\\
&&\hspace*{3.7cm}\Biggl(1+\frac{m_3^2}{2 m_1^2}\,
\frac{1}{(1-x)^2\,(1+x)^2\,(5x^3-22x^2+41x+8)^2}\,
(169x^{10}-\Biggr.\\
&&\hspace*{3.7cm}
1306x^9+\Biggl.5334x^8-13168x^7+22638x^6-25436x^5+10336x^4+\Biggr.\\
&&\hspace*{3.7cm}14672x^3-
8271x^2\Biggl.-1514x+642)\Biggr)+{\cal
O}\left(\frac{m_3^4}{m_1^4}\right),\\[5mm]
&&\frac{\Gamma(B_c\to c\bar c[{\scriptstyle ^3D_2}]\rho)}
{\Gamma(B_c\to c\bar c[{\scriptstyle ^3D_2}]\pi)}=B_{c\bar c[{\scriptstyle
^3D_2}]}\, \Biggl(1+\frac{m_3^2}{24 m_1^2}\,
\frac{1} {(1-x)^6\,(1+x)^2}
(13x^8-96x^7+433x^6-\\&&\hspace*{3.7cm}822x^5+1201x^4-124x^3-873x^2+1230x+122)
\Biggr)+{\cal O}\left(\frac{m_3^4}{m_1^4}\right),\\[3mm]
&&\frac{\Gamma(B_c\to c\bar c[{\scriptstyle ^3D_3}]\rho)}
{\Gamma(B_c\to c\bar c[{\scriptstyle ^3D_3}]\pi)}=
B_{c\bar c[{\scriptstyle ^3D_3}]}\, \Biggl(1+\frac{m_3^2}{12 m_1^2}\,
\frac{1}{(1-x)^2\,(1+x)^2}(x^6-8x^5+52x^4-\\&&\hspace*{3.7cm}96x^3-15x^2+152x-2
2)\Biggr)+{\cal O}\left(\frac{m_3^4}{m_1^4}\right).
\end{eqnarray*}
\normalsize
The factors $B$ for various charmonium states tend to the ratio of leptonic
constants of $\rho$ and $\pi$ mesons, if we neglect the $\rho$ meson mass,
while in general case we obtain \small
\begin{eqnarray*}
&&B_{\eta_c}=\frac{f_\rho^2}{f_\pi^2}\,
\frac{x^6\,(1-x)^5}
{(1+x)\left((1-x)^2-\displaystyle\frac{m_3^2}{m_1^2}\right)^3}
\,\sqrt{(1-x)^2\,(1+x)^2-\frac{2m_3^2}{m_1^2}\,(1+x^2)+\frac{m_3^4}{m_1^4}},
\\[3mm]
&&B_{J/\psi}=B_{\eta_c}\,\frac{x^6\,(1-x)^2}{\left((1-x)^2-\displaystyle\frac{m
_3^2}
{m_1^2}\right)},\\[3mm]
&&B_{h_c}=\frac{f_\rho^2}{f_\pi^2}\,
\frac{(1-x)^7}{(1+x)\left((1-x)^2-\displaystyle\frac{m_3^2}{m_1^2}\right)^4}
\,\sqrt{(1-x)^2\,(1+x)^2-\frac{2m_3^2}{m_1^2}\,(1+x^2)+\frac{m_3^4}{m_1^4}},
\\[3mm]
&&B_{\chi_{c0}}=B_{h_c}\,\frac{(1-x)^2}{\left((1-x)^2-\displaystyle\frac{m_3^2}
{m_1^2}\right)},\\[3mm]
&&B_{\chi_{c1}}=B_{\chi_{c0}},\\[3mm]
&&B_{\chi_{c2}}=\frac{f_\rho^2}{f_\pi^2}\,
\frac{(1-x)^9}{(1+x)\left((1-x)^2-\displaystyle\frac{m_3^2}{m_1^2}\right)^5}
\,\frac{1}{\sqrt{\displaystyle{{(1-x)^2\,(1+x)^2-\frac{2m_3^2}{m_1^2}\,(1+x^2)+
\frac{m_3^4}{m_1^4}}}}},\\[3mm]
&&B_{c\bar c[{\scriptstyle ^1D_2}]}=\frac{f_\rho^2}{f_\pi^2}\,
\frac{(1-x)^9\,(1+x)}{\left((1-x)^2-\displaystyle\frac{m_3^2}{m_1^2}\right)^4}
\,\frac{1}{\sqrt{\displaystyle{{(1-x)^2\,(1+x)^2-\frac{2m_3^2}{m_1^2}\,(1+x^2)+
\frac{m_3^4}{m_1^4}}}}},\\[3mm]
&&B_{c\bar c[{\scriptstyle ^3D_1}]}=\frac{f_\rho^2}{f_\pi^2}
\frac{(1-x)^{11}}{(1+x)\left((1-x)^2-\displaystyle\frac{m_3^2}{m_1^2}\right)^6}
\,\sqrt{(1-x)^2\,(1+x)^2-\frac{2m_3^2}{m_1^2}\,(1+x^2)+\frac{m_3^4}{m_1^4}},
\\[3mm]
&&B_{c\bar c[{\scriptstyle ^3D_2}]}=
B_{c\bar c[{\scriptstyle ^3D_3}]}=
B_{c\bar c[{\scriptstyle ^3D_1}]}.
\end{eqnarray*}
\normalsize
Finally, we note that the most significant numerical correction due to the
nonzero $\rho$ meson mass comes from the phase space in the coefficients $B$.


\begin{thebibliography}{999}
\bibitem{cdf}
F. Abe et al., CDF Collaboration, Phys. Rev. Lett. {\bf 81}, 2432 (1998),
Phys. Rev. {\bf D58}, 112004 (1998).
\bibitem{QCDSRBc}
P.Colangelo, G.Nardulli, N.Paver, Z.Phys. {\bf C57}, 43 (1993);\\
E.Bagan et al., Z. Phys. {\bf C64}, 57 (1994);\\
V.V.Kiselev, A.V.Tkabladze, Phys. Rev. {\bf D48}, 5208 (1993);\\
V.V.Kiselev, A.K.Likhoded, O.A.Onishchenko, Nucl. Phys. {\bf B569}, 473
(2000);\\
V.V.Kiselev, A.K.Likhoded, A.E.Kovalsky, Nucl. Phys. {\bf B585}, 353 (2000),
hep-ph/0006104 (2000).
\bibitem{PMBc}
M.Lusignoli, M.Masetti, Z. Phys. {\bf C51}, 549 (1991);\\
V.V.Kiselev, Mod. Phys. Lett. {\bf A10}, 1049 (1995);\\
V.V.Kiselev, Int. J. Mod. Phys. {\bf A9}, 4987 (1994);\\
V.V.Kiselev, A.K.Likhoded, A.V.Tkabladze, Phys. Atom. Nucl.  {\bf 56}, 643
(1993), Yad. Fiz. {\bf 56}, 128 (1993);\\
V.V.Kiselev, A.V.Tkabladze, Yad. Fiz. {\bf 48}, 536 (1988);\\
S.S.Gershtein et al., Sov. J. Nucl. Phys. {bf 48}, 327 (1988), Yad. Fiz. {\bf
48}, 515 (1988);\\
G.R.Jibuti, Sh.M.Esakia, Yad. Fiz. {\bf 50}, 1065 (1989), 
Yad. Fiz. {\bf 51}, 1681 (1990);\\
D.Scora, N.Isgur, Phys. Rev. {\bf D52}, 2783 (1995);\\
A.Yu.Anisimov, I.M.Narodetskii, C.Semay, B.Silvestre--Brac, Phys. Lett. {\bf
B452}, 129 (1999);\\  
A.Yu.Anisimov, P.Yu.Kulikov, I.M.Narodetsky, K.A.Ter-Martirosian, 
Phys. Atom. Nucl. {\bf 62}, 1739 (1999), Yad. Fiz. {\bf 62}, 1868 (1999);\\
M.A.Ivanov, J.G.Korner, P.Santorelli, Phys. Rev. {\bf D63}, 074010 (2001);\\
P.Colangelo, F.De Fazio, Phys. Rev. {\bf D61}, 034012 (2000).
\bibitem{OPEBc}
I.Bigi, Phys. Lett. {\bf B371}, 105 (1996);\\
M.Beneke, G.Buchalla, {Phys. Rev.} {\bf D53}, 4991 (1996);\\
A.I.Onishchenko, [hep-ph/9912424];\\
Ch.-H.Chang, Sh.-L.Chen, T.-F.Feng, X.-Q.Li, Commun. Theor. Phys. {\bf 35}, 51
(2001), 
Phys. Rev. {\bf D64}, 014003 (2001).
\bibitem{NRQCD}
G.T.Bodwin, E.Braaten, G.P.Lepage, Phys. Rev. {\bf D51}, 1125 (1995)
[Erratum-ibid.  {\bf D55}, 5853 (1995)];\\
T.Mannel, G.A.Schuler, Z. Phys. {\bf C67}, 159 (1995).
\bibitem{Brodsky}
G.P.Lepage, S.J.Brodsky, Phys. Rev. {\bf D23}, 2157 (1980).
\bibitem{HSBc}
S.S.Gershtein et al., preprint IHEP 98-22 (1998) [hep-ph/9803433];\\
V.V.Kiselev, Phys.Lett. {\bf B372}, 326 (1996), hep-ph/9605451;\\
O.N.Pakhomova, V.A.Saleev, Phys. Atom. Nucl. {\bf 63}, 1999 (2000); Yad.Fiz.
{\bf 63}, 2091 (2000) [hep-ph/9911313];\\
V.A.Saleev, Yad. Fiz. 64 (2001) (in press) [hep-ph/0007352].
\bibitem{Anisovich}
V.V.Anisovich, D.I.Melikhov, V.A.Nikonov, Phys. Rev. {\bf D55}, 2918 (1997),
{\bf D52}, 5295 (1995),
Phys.\ Atom.\ Nucl.\  {\bf 57},  490 (1994)
[Yad.\ Fiz.\  {\bf 57}, 520 (1994)].
\bibitem{fact}
M.Dugan and B.Grinstein, Phys. Lett. {\bf B255}, 583 (1991);\\
M.A.Shifman, Nucl. Phys. {\bf B388}, 346 (1992);\\
B.Blok, M.Shifman, Nucl. Phys. {\bf B389}, 534 (1993).
\bibitem{n4}
B.Guberina et al., { Nucl. Phys.} {\bf B174}, 317 (1980);\\
M.B.Voloshin, M.A.Shifman, { Yad. Fiz.} {\bf 41}, 187 (1985);\\
M.B.Voloshin, M.A.Shifman, { Sov.\ Phys.\ JETP } {\bf 64}, 698 (1986).
\bibitem{n5}
L.Bergstr\"om et  al.,  {Phys.  Rev.} {\bf  D43}, {2157} (1991).
\bibitem{EichQuigg}
E.Eichten, C.Quigg, Phys. Rev. {\bf D49}, 5845 (1994);\\
S.S.Gershtein et al., Phys. Rev. {\bf D51}, 3613 (1995);\\
S.S.Gershtein et al., {Usp. Fiz. Nauk} {\bf 165}, {3} (1995);\\
E.Eichten, C.Quigg, Phys. Rev. {\bf D52}, 1726 (1995);\\
V.V.Kiselev, Phys.\ Part.\ Nucl.\  {\bf 31}, 538 (2000)
[Fiz.\ Elem.\ Chast.\ Atom.\ Yadra {\bf 31}, 1080 (2000)];\\
V.V.Kiselev, Int.\ J.\ Mod.\ Phys.\  {\bf A11}, 3689 (1996);\\
V.V.Kiselev, Nucl.\ Phys.\ {\bf B406}, 340 (1993).
\bibitem{chache}
Ch.-H.Chang, Y.-Q.Chen, G.-L.Wang, H.-Sh.Zong, hep-ph/0103036 (2001).
\bibitem{n6}
D.E.Groom et al., {Eur. Phys. J.} {\bf C15}, 1 (2000).
\bibitem{10}
C.-H.Chang, Y.-Q.Chen, Phys. Rev. {\bf D49}, 3399 (1994).
\bibitem{Vary}
A. Abd El-Hady, J.H.Munoz, J.P. Vary, Phys. Rev. {\bf D62}, 014019 (2000).
\end{thebibliography}
\end{document}